\documentclass[11pt,a4paper]{article}
\usepackage{tikz}
\tikzstyle{hybb}=[rectangle,draw,minimum size=3.5mm]
\tikzstyle{tree}=[circle,draw,minimum size=3.75mm]
\newcommand{\etq}[1]{\draw (#1) node {$#1$}; }
\usepackage{amssymb}

\renewcommand{\le}{\leqslant}

\usepackage{url}

\begin{document}

\title{A Perl Package and an Alignment Tool for Phylogenetic Networks}
\author{\textbf{Gabriel Cardona}\\
Department of Mathematics\\
and Computer Science\\
University of the Balearic Islands\\
E-07122 Palma de Mallorca\\
Spain
\and
\textbf{Francesc Rossell\'o}\\
Research Institute of Health Science\\
University of the Balearic Islands\\
E-07122 Palma de Mallorca\\
Spain
\and \textbf{Gabriel Valiente}\\
Algorithms, Bioinformatics, Complexity\\
and Formal Methods Research Group\\
Technical University of Catalonia\\
E-08034 Barcelona\\
Spain
}
\maketitle

\begin{abstract}
Phylogenetic networks are a generalization of phylogenetic trees that allow for the representation of evolutionary events acting at the population level, like recombination between genes, hybridization between lineages, and lateral gene transfer. While most phylogenetics tools implement a wide range of algorithms on phylogenetic trees, there exist only a few applications to work with phylogenetic networks, and there are no open-source libraries either.
In order to improve this situation, we have developed a Perl package that relies on the BioPerl bundle and implements many algorithms on phylogenetic networks. We have also developed a Java applet that makes use of the aforementioned Perl package and allows the user to make simple experiments with phylogenetic networks without having to develop a program or Perl script by herself.
The Perl package has been accepted as part of the BioPerl bundle. It can be downloaded  from the url \texttt{http://dmi.uib.es/\~{}gcardona/BioInfo/Bio-Phylo\allowbreak Network.tgz}. The web-based application is available at the url \texttt{http://dmi.uib.es/\allowbreak\~{}gcardona/BioInfo/}. The Perl package includes full documentation of all its features.
\end{abstract}

\section*{Background}
We briefly recall some definitions and results from~\cite{cardona.ea:tcbb:2007} on phylogenetic networks. 

A \emph{phylogenetic network} on a set $S$ of taxa is any rooted directed acyclic graph whose leaves (those nodes without outgoing edges) are bijectively labeled by the set $S$. 

Let $N=(V,E)$ be a phylogenetic network on $S$. A node $u\in V$ is said to be a \emph{tree node} if it has, at most, one incoming edge; otherwise it is called a \emph{hybrid node}. A phylogenetic network on $S$ is a \emph{tree-child phylogenetic network} if every node either is a leaf or has at least one child that is a tree node.

Let $S=\{\ell_1,\ldots,\ell_n\}$ be the set of leaves. We define the \emph{$\mu$-vector} of a node $u\in V$ as the vector $\mu(u)=(m_1(u),\ldots, m_n(u))$, where $m_i(u)$ is the number of 
different paths from $u$ to the leaf $\ell_i$. The multiset $\mu(N)=\{\mu(v)\mid v\in V\}$ is called the \emph{$\mu$-representation} of $N$ and, provided that $N$ is a tree-child phylogenetic network, it turns out to completely characterize $N$, up to isomorphisms, among all tree-child phylogenetic networks on $S$.

This allows us to define a distance on the set of tree-child phylogenetic networks on $S$: the \emph{$\mu$-distance} between two given networks $N_1$ and $N_2$ is the symmetric difference of their $\mu$-representations, 
\[
d_\mu(N_1,N_2)=|\mu(N_1)\mathop{\triangle} \mu(N_2)|\,.
\]
This defines a true distance, and when $N_1$ and $N_2$ are phylogenetic trees, it coincides with the well-known partition distance~\cite{robinson.foulds:mb81}.

This representation also allows us to define an optimal alignment between two tree-child phylogenetic networks on $S$, say $n=|S|$. Given two such networks $N_1=(V_1,E_1)$ and $N_2=(V_2,E_2)$ (where, for the sake of simplicity, we assume $|V_1|\le|V_2|$), an \emph{alignment} is just an injective mapping $M:V_1\to V_2$. The \emph{weight} of this alignment is 
\[
w(M)=\sum_{v\in V_1}\big(\|\mu(v)-\mu(M(v))\|+\chi(v,M(v))\big),
\]
where $\|\cdot\|$ stands for the Manhattan norm of a vector and $\chi(u,v)$ is $0$ if both $u$ and $v$ are tree nodes or hybrid nodes, and $1/(2n)$ if one of them is a tree node and the other one is a hybrid node. An \emph{optimal alignment} is, then, an alignment with minimal weight.

\section*{The Extended Newick Format}

The eNewick (for ``extended Newick'') string defining a phylogenetic network appeared in the packages \textsc{PhyloNet} \cite{PhyloNet} and \textsc{NetGen} \cite{NetGen} related to phylogenetic networks, with some differences between them. The former encodes a phylogenetic network with $k$ hybrid nodes as a series of $k$ trees in Newick format, while the latter encodes it as a single tree in Newick format but with $k$ repeated nodes.

Whereas the Perl module we introduce here accepts both formats as input, a complete standard for eNewick is implemented, based mainly on \textsc{NetGen} and following the suggestions of D.~Huson and M. M.~Morin (among others), to make it as complete as possible. The adopted standard has the practical advantage of encoding a whole phylogenetic network as a single string, and it also includes mandatory tags to distinguish among the various hybrid nodes in the network.

The procedure to obtain the eNewick string representing a phylogenetic network $N$ goes as follows: Let $\{H_1,\dots,H_m\}$ be the set of hybrid nodes of $N$, ordered in any fixed way. For each hybrid node $H=H_i$, say with parents $u_1,u_2,\ldots,u_k$ and children $v_1,v_2,\ldots,v_\ell$: split $H$ in $k$ different nodes; let the first copy be a child of $u_1$ and have all $v_1,v_2,\ldots,v_\ell$ as its children; let the other copies be children of $u_2,\ldots,u_k$ (one for each) and have no children. Label each of the copies of $H$ as
\begin{center}
\verb|[label]#[type]tag[:branch_length]|
\end{center}
where the parameters are:
\begin{itemize}
\item  \verb|label| (optional) string providing a labelling for
the node;
\item \verb|type| (optional) string indicating if the node $H$ corresponds to a hybridization (indicated by \verb|H|) or a lateral gene transfer (indicated by \verb|LGT|) event; note that other types can be considered in the future;
\item \verb|tag| (mandatory) integer $i$ identifying the node $H=H_i$.
\item \verb|branch_length|  (optional) number giving the length of the branch from the copy of $H$ under consideration to its parent.
\end{itemize}

In this way, we get a tree whose set of leaves is the set of leaves of the original network together with the set of hybrid nodes (possibly repeated). Then, the Newick string of the obtained tree (note that some internal nodes will be labeled and some leaves will be repeated) is the eNewick string of the phylogenetic network. The leftmost occurrence of each hybrid node in an eNewick string corresponds to the full description of the network rooted at that node, and although node labels are optional, all labeled occurrences of a hybrid node in an eNewick string must carry the same label.

\begin{figure}
\begin{center}
\begin{tikzpicture}[thick,>=stealth,xscale=0.75,yscale=0.5]
\draw (2,0) node[tree] (r) {}; \etq r
\draw (0.7,-2) node[tree] (x) {}; \etq x
\draw (2,-2.5) node[hybb] (h) {}; \etq h
\draw (3.3,-2) node[tree] (y) {}; \etq y
\draw (0.5,-4) node[tree] (1) {}; \etq 1
\draw (2,-4) node[tree] (2) {}; \etq 2
\draw (3.5,-4) node[tree] (3) {}; \etq 3
\draw[->] (r)--(x);
\draw[->] (r)--(y);
\draw[->] (x)--(h);
\draw[->] (x)--(1);
\draw[->] (h)--(2);
\draw[->] (y)--(h);
\draw[->] (y)--(3);
\end{tikzpicture}
\hfil
\begin{tikzpicture}[thick,>=stealth,xscale=0.75,yscale=0.5]
\draw (2,0) node[tree] (r) {}; \etq r
\draw (0.7,-2) node[tree] (x) {}; \etq x
\draw (1.7,-2.5) node[hybb] (A) {}; \draw (A) node {$h$};
\draw (2.3,-2.5) node[hybb] (h) {}; \etq h
\draw (3.3,-2) node[tree] (y) {}; \etq y
\draw (0.5,-4) node[tree] (1) {}; \etq 1
\draw (1.7,-4) node[tree] (2) {}; \etq 2
\draw (3.5,-4) node[tree] (3) {}; \etq 3
\draw[->] (r)--(x);
\draw[->] (r)--(y);
\draw[->] (x)--(A);
\draw[->] (x)--(1);
\draw[->] (A)--(2);
\draw[->] (y)--(h);
\draw[->] (y)--(3);
\end{tikzpicture}
\end{center}
\caption{\label{forest}A phylogenetic network $N$ (left), and  tree (right) associated to $N$ for computing its eNewick string.} 
\end{figure}
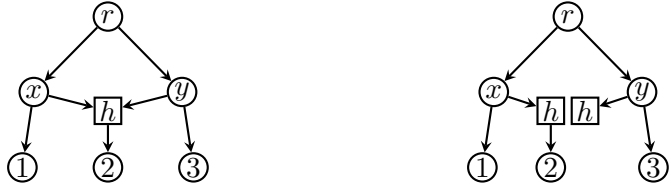

Consider, for example, the phylogenetic network depicted together with its decomposition in Figure~\ref{forest}.
The eNewick string for this network would be
\verb|((1,(2)#H1),(#H1,3));|
or
\verb|((1,(2)h#H1)x,(h#H1,3)y)r;|
if all internal nodes are labeled. The leftmost occurrence of the hybrid node in the latter string corresponds to the full description of the network rooted at that node: \verb|(2)h#H1|.

Obviously, the procedure to recover a network from its eNewick string is as simple as recovering the tree and identifying those nodes that are labeled as hybrid nodes with the same identifier.

\begin{figure}
\begin{center}
\begin{tikzpicture}[thick,>=stealth,xscale=0.75,yscale=0.5]
\draw (2,0) node[tree] (r) {}; \etq r
\draw (1.25,-2) node[tree] (x) {}; \etq x
\draw (0.5,-4) node[tree] (1) {}; \etq 1
\draw (2,-4) node[tree] (2) {}; \etq 2
\draw (3.5,-4) node[tree] (3) {}; \etq 3
\draw[->] (r)--(x);
\draw[->] (x)--(1);
\draw[->] (x)--(2);
\draw[->] (r)--(3);
\draw[->] (intersection of 1,-2.75--3,-2.75 and x--2)--(intersection of 1,-2.75--3,-2.75 and r--3);
\end{tikzpicture}
\hfil
\begin{tikzpicture}[thick,>=stealth,xscale=0.75,yscale=0.5]
\draw (2,0) node[tree] (r) {}; \etq r
\draw (1.25,-2) node[tree] (x) {}; \etq x
\draw (intersection of 1,-3--3,-3 and x--2) node[tree] (y) {}; \etq y
\draw (intersection of 1,-2.5--3,-2.5 and r--3) node[hybb] (h) {}; \etq h
\draw (0.5,-4) node[tree] (1) {}; \etq 1
\draw (2,-4) node[tree] (2) {}; \etq 2
\draw (3.5,-4) node[tree] (3) {}; \etq 3
\draw[->] (r)--(x);
\draw[->] (r)--(h);
\draw[->] (x)--(y);
\draw[->] (x)--(1);
\draw[->] (y)--(2);
\draw[->] (y)--(h);
\draw[->] (h)--(3);
\end{tikzpicture}
\end{center}
\caption{\label{LGT}Representation of a lateral gene transfer event (left) as a hybrid node in a phylogenetic network (right).}
\end{figure}
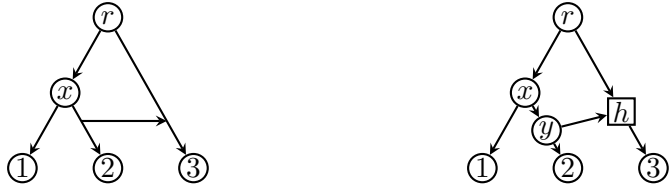

Notice that gene transfer events can be represented in a unique way as hybrid nodes. Consider, for example, the lateral gene transfer event depicted in Figure~\ref{LGT}, where a gene is transferred from species 2 to species 3 after the divergence of species 1 from species 2. The eNewick string \verb|((1,(2,(3)h#LGT1)y)x,h#LGT1)r;| describes such a phylogenetic network. A program interpreting the eNewick string can use the information on node types in different ways; for instance, to render tree nodes circled, hybridization nodes boxed, and lateral gene transfer nodes as arrows between edges.

\section*{The Perl Module}

The Perl module \texttt{Bio::PhyloNetwork} implements all the data structures needed to work with tree-child phylogenetic networks, as well as algorithms for:
\begin{itemize}
\item reconstructing a network from its eNewick string (in all its different flavours),
\item reconstructing a network from its $\mu$-representation,
\item exploding a network into the set of its induced subtrees,
\item computing the $\mu$-representation of a network and the $\mu$-distance between two networks,
\item computing an optimal alignment between two networks,
\item computing tripartitions \cite{moret.ea:2004,cardona.ea:07a} and the tripartition error between two networks, and
\item testing if a network is time consistent~\cite{baroni.ea:sb06}, and in such a case, computing a temporal representation.
\end{itemize}

The underlying data structure is a \texttt{Graph::Directed} object, with some extra data, for instance the $\mu$-representation of the network. It makes use of the Perl module \texttt{Bio::Phylo\allowbreak Network::muVector} that implements basic arithmetic operations on $\mu$-vectors. Two extra modules, \texttt{Bio::PhyloNetwork::Factory} and \texttt{Bio::PhyloNetwork::RandomFactory}, are provided for the sequential and random generation (respectively) of all tree-child phylogenetic networks on a given set of taxa.

\section*{The web interface and the java applet}

The web interface, available at~\texttt{http://dmi.uib.es/\~{}gcardona/BioInfo/}, allows the user to input one or two phylogenetic networks, given by their eNewick strings. A Perl script processes these strings and uses the \texttt{Bio::PhyloNetwork} package to compute all available data for them, including a plot of the networks that can be downloaded in PS format; these plots are generated through the application \texttt{GraphViz} and its companion Perl package.

Given two networks on the same set of leaves, their $\mu$-distance is also computed, as well as an optimal alignment between them. The algorithm to compute such an alignment relies on the Hungarian algorithm~\cite{munkres:57}. If their sets of leaves are not the same, their \emph{topological restriction} on the set of common leaves is first computed followed by the $\mu$-distance and an optimal alignment.

A Java applet displays the networks side by side, and whenever a node is selected, the corresponding node in the other network (with respect to the optimal alignment) is highlighted, provided it exists. This is also extended to edges. Similarities between the networks are thus evident at a glance and, since the weight of each matched node is also shown, it is easy to see where the differences are.

\section*{Authors contributions}

All authors conceived the method, prepared the manuscript, contributed to the discussion, and have approved the final manuscript.
GC implemented the software.
GV also implemented part of the software.
    
\section*{Acknowledgements}

The research described in this paper has been partially supported by the Spanish CICYT project TIN 2004-07925-C03-01 GRAMMARS and by Spanish DGI projects MTM2006-07773 COMGRIO and MTM2006-15038-C02-01. 

%\bibliographystyle{plain}
%\bibliography{aligner-arxiv}

\end{document}